\pdfoutput=1

\documentclass[final,3p,times,twocolumn]{elsarticle}

\usepackage{graphicx}

\usepackage{amsmath}

\usepackage{lineno}

\journal{NIM A Proceedings of RICAP 2011}

\begin{document}

\begin{frontmatter}



\title{Particle Physics in Ice with IceCube DeepCore}


\author{Tyce DeYoung\\
{\small for the IceCube collaboration}}

\address{Department of Physics, Pennsylvania State University,
  University Park, PA 16802, U.S.A.\\
 E-mail: deyoung@psu.edu}

\begin{abstract}
  The IceCube Neutrino Observatory is the world's largest high energy
  neutrino telescope, using the Antarctic ice cap as a Cherenkov
  detector medium. DeepCore, the low energy extension to IceCube, is
  an infill array with a fiducial volume of around 30 MTon in the
  deepest, clearest ice, aiming for an energy threshold as low as 10
  GeV and extending IceCube's sensitivity to indirect dark matter
  searches and atmospheric neutrino oscillation physics. We will
  discuss the analysis of the first year of DeepCore data, as well as
  ideas for a further extension of the particle physics program in the
  ice with a future PINGU detector.
\end{abstract}

\begin{keyword}
astroparticle physics; neutrino oscillations; dark matter
\end{keyword}

\end{frontmatter}

\section{Introduction}\label{sec:intro}

The IceCube neutrino telescope, now fully operational at depths of
1450-2450 m below the surface of the Antarctic ice cap, was designed
to detect high energy neutrinos from astrophysical accelerators of
cosmic rays.  Although the energy threshold of a large volume neutrino
detector is not a sharp function, the original IceCube design focused
on efficiency for neutrinos at TeV energies and above.  Recently, the
IceCube collaboration decided to augment the response of the detector
at lower energies with the addition of DeepCore, a fully contained
subarray aimed at improving the sensitivity of IceCube to neutrinos
with energies in the range of 10's of GeV to a few hundred GeV.  This
energy range is of interest for several topics related to particle
physics, including measurements of neutrino oscillations and searches
for neutrinos produced in the annihilation or decay of dark matter.

DeepCore consists of an additional eight strings of photosensors
(Digital Optical Modules, or DOMs) comprising 10'' Hamamatsu
photomultiplier tubes and associated data acquisition electronics
housed in standard IceCube glass pressure vessels.  For most of the
DeepCore DOMs, the standard IceCube R7081 PMTs were replaced with
7081MOD PMTs with Hamamatsu's new super-bialkali photocathode.  These
PMTs provide approximately 35\% higher quantum efficiency (averaged
over the detected Cherenkov spectrum) than the standard bialkali PMTs.

\begin{figure}
  \begin{center}
  \includegraphics[width=0.8\columnwidth]{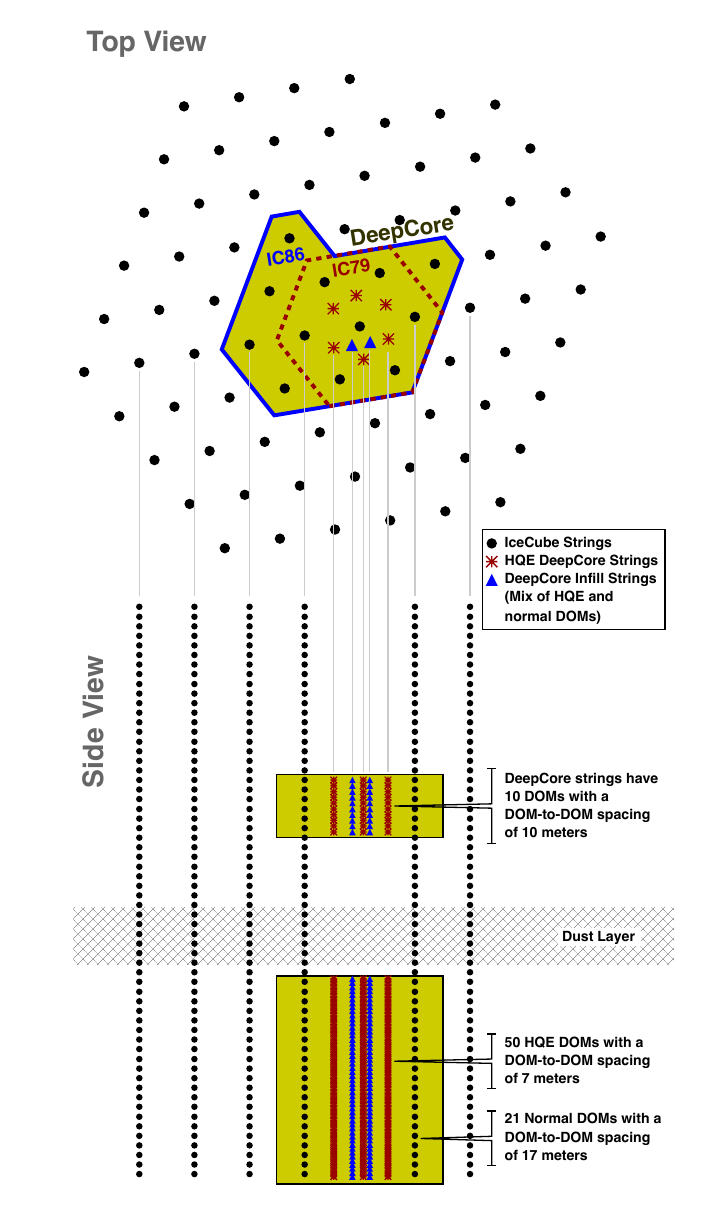}
  \caption{Schematic layout of DeepCore within IceCube.  The shaded
    region indicates the fiducal volume of DeepCore, at the bottom
    center of IceCube, plus the extra veto cap of DOMs deployed at
    shallower depths to reinforce the veto against
    vertically-downgoing atmospheric muons.  This schematic depicts
    both the DeepCore configuration used in 2010, when 79 IceCube
    strings were operational, and the final DeepCore layout and fiducial region used in the 2011
    run.  
    \label{fig:layout}}
  \end{center}
\end{figure}

Sited at the bottom center of the IceCube array, DeepCore benefits
from the high optical quality of the ice at depths of 2100-2450 m,
with an attenuation length of approximately 50 m in the blue
wavelengths at which most Cherenkov photons are detected in ice.
DeepCore also benefits from the ability of the standard IceCube
sensors to detect atmospheric muons penetrating the ice from cosmic
ray air showers above the detector, allowing substantial reduction in
the background rate by vetoing events where traces of penetrating
muons are seen.  Each DeepCore string bears 50 DOMs in the fiducial
region, with an additional 10 DOMs deployed at shallower depths to
improve the vetoing efficiency for steeply vertical muons.  In
addition to the new DeepCore strings, the DeepCore fiducial volume for
analysis includes 12 standard IceCube strings, chosen so that the
fiducial region is shielded on all sides by a veto region consisting
of three rows of standard IceCube strings, as shown in
Fig.~\ref{fig:layout}. 

The random noise rate of IceCube DOMs is quite low (around 500 Hz, on
average) due to the low temperatures and radiopurity of the ice cap.
This permits DeepCore to be operated with a very low trigger
threshold, demanding that 3 DOMs within the DeepCore fiducial region
detect light in ``local coincidence'' within a period of no more than
2500 ns.  The local coincidence criterion counts DOMs as being hit (i.e., having
detected light) only if one of the four neighboring DOMs on a string
(two above and two below) also registers a hit within $\pm1 \;\mu$s.
Most of the resulting 185 Hz of triggers are due to stray light from
muons which simultaneously satisfy the main IceCube trigger condition
of 8 DOMs hit in local coincidence within 5 $\mu$s, but the DeepCore
trigger contributes an additional (exclusive) rate of around 10 Hz.

The vast majority of the events which trigger DeepCore, irrespective
of whether they also trigger IceCube, are due to either penetrating
atmospheric muons or random coincidences of dark noise.  Immediately
after data acquisition, events triggering DeepCore are subjected to an
online data rejection algorithm which calculates a characteristic time
and location for the activity observed in the DeepCore fiducial
region, as an initial estimate of the putative neutrino vertex.  The
estimated location is the average position of the hit DOMs, and the
time is determined by subtracting the time of flight $dn/c$ of an unscattered
photon emitted from that location from the observed arrival time of
the first photon to hit each DOM.  After outliers due to dark noise or
scattered light are removed, the
average inferred emission time is used as the estimated
time of the underlying physics event.

\begin{figure}
  \includegraphics[width=\columnwidth]{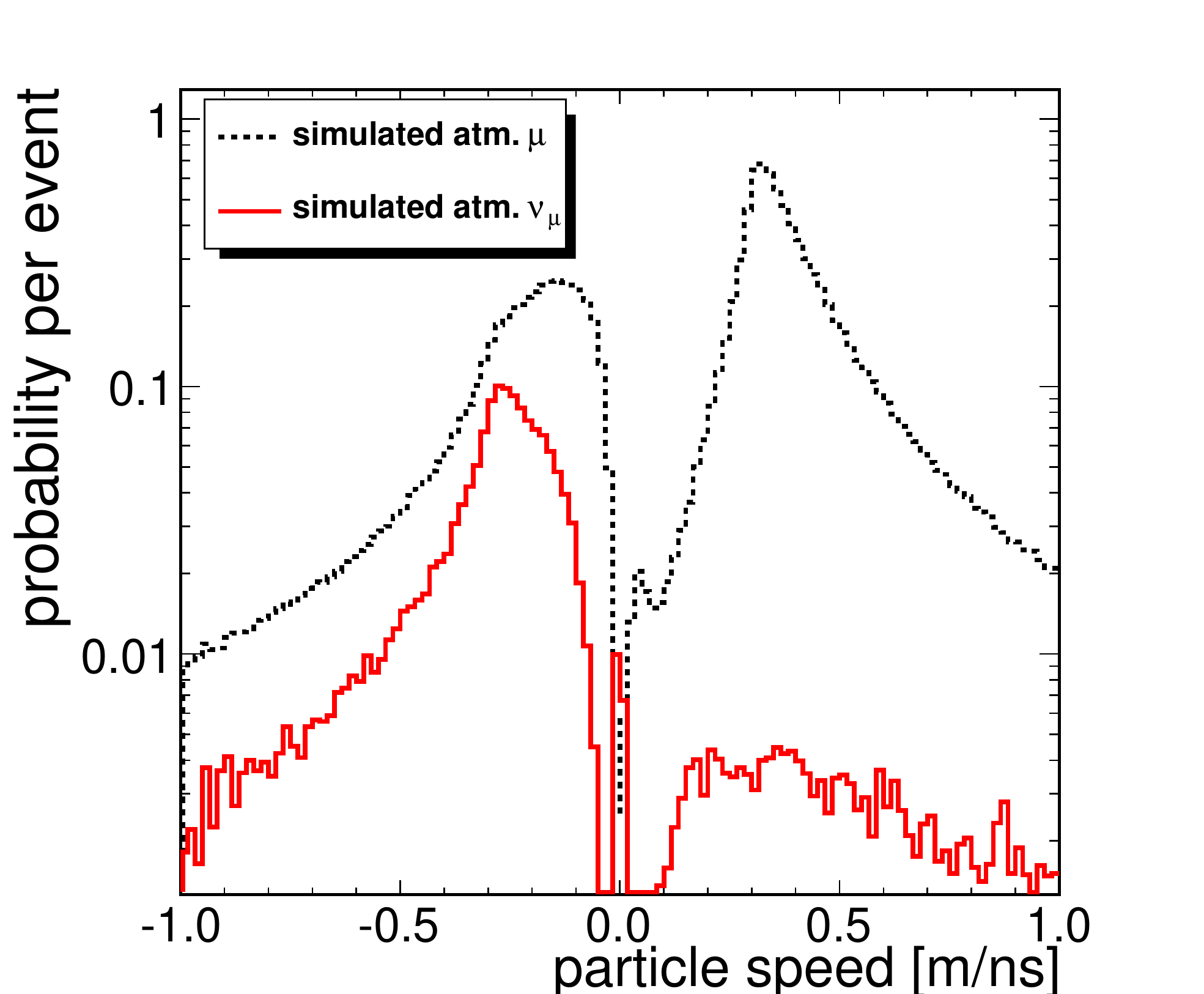}
  \caption{Distribution of probabilities of observing hits leading to
    a given inferred particle speed, for simulated
    atmospheric muons (dashed line) and atmospheric neutrinos (solid).
    Positive speeds indicate activity in the veto region prior to that
    in the DeepCore volume, and a peak around $c =$ 0.3 m/ns is
    visible for penetrating muons.  The integral of each distribution
    corresponds to the mean number of hits observed in the veto region
    for the given class of events.
    \label{fig:veto}}
\end{figure}

Based on this estimated time and location, every locally coincident
hit recorded in the veto region prior to the vertex time is examined
to determine whether it lies on the light cone connecting it with the
estimated event vertex.  The distributions of the inferred speed
required to connect hits in the veto region to the DeepCore vertex,
for both simulated atmospheric muons and simulated neutrinos, is shown
in Fig.~\ref{fig:veto}; positive speeds indicate hits occuring in the
veto region prior to the DeepCore vertex time.  If any hits are found
with inferred speeds between +0.25 and +0.4 m/ns, the event is
rejected as being most likely due to an atmospheric muon.  This
algorithm reduces the event rate by more than two orders of magnitude,
to 18 Hz, while retaining over 99\% of simulated triggered events due to
neutrinos interacting within the fiducial volume.  Additional
background rejection criteria are applied offline, depending on the
goals of each physics analysis making use of these data.

\begin{figure}
  \includegraphics[width=\columnwidth]{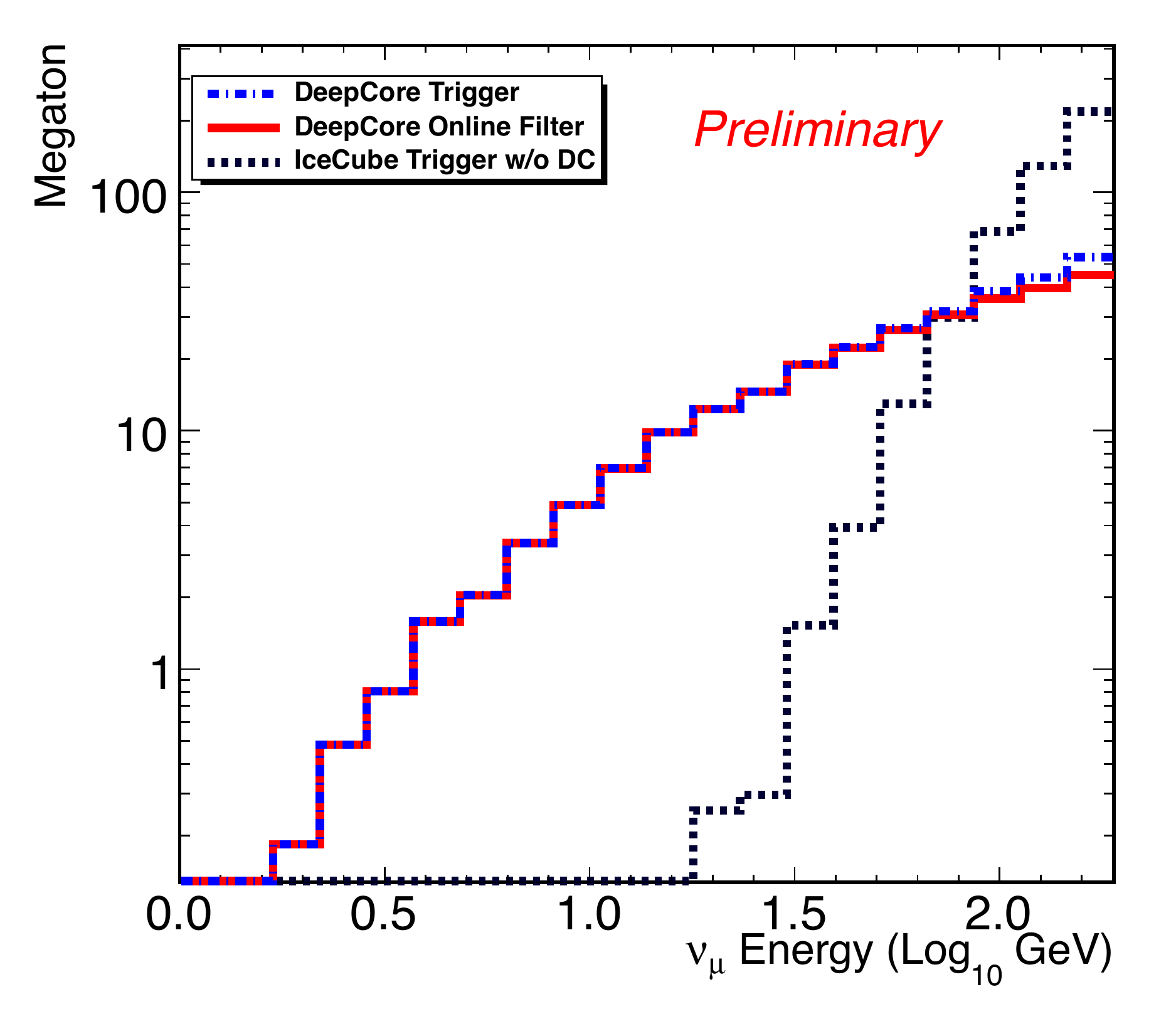}
  \caption{Effective volume of DeepCore for muon neutrinos at trigger
    level (solid) and after application of the online veto algorithm
    described in the text (dot-dashed line).  The effective volume of
    IceCube as originally proposed is shown for comparison.  
    \label{fig:nuMuVolume}}
\end{figure}

The effective volume of the DeepCore detector for detection low energy
muon neutrinos, accounting for this online data filter, is shown in
Fig.~\ref{fig:nuMuVolume}.  It should be stressed that this effective
volume curve does \emph{not} include losses due to later background
rejection or event quality criteria.  The contribution of DeepCore to
low energy analysis is evident in the fact that despite its relatively
small geometric volume, around 3\% that of IceCube, the overall sample
of neutrino events below 100 GeV consists primarily of those detected
by DeepCore.  This energy range is of considerable interest for
several topics in particle physics, including searches for dark matter
and measurements of neutrino oscillations.  While DeepCore does not
have a sharp energy threshold, it retains around 7 megatons of
effective volume at energies as low as 10 GeV.  Further details
regarding DeepCore's instrumentation and performance are available in
Ref.~\cite{Collaboration:2011ym}.

\section{Observation of Neutrino-Induced Cascades}\label{sec:cascades}

\begin{figure}
  \includegraphics[width=\columnwidth]{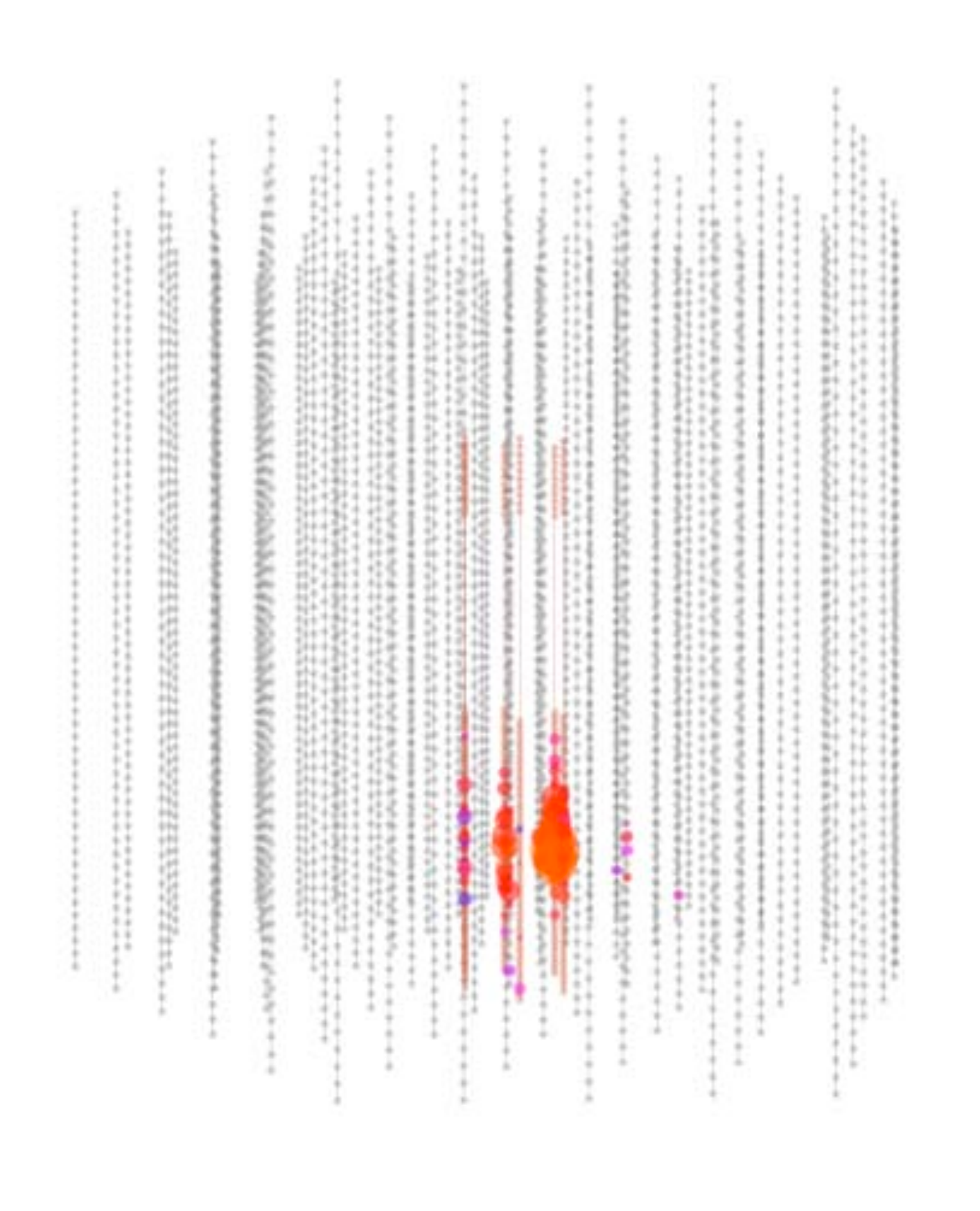}
  \caption{Candidate neutrino-induced cascade observed in DeepCore in
    the 2010 data run.  Each black dot indicates a  DOM.  Colored dots
    represent DOMs that detected light during the event, with the size
    of the dot proportional to the amount of light detected.  The
    color indicates the relative arrival time of the first photon
    detected by that DOM, running through the spectrum from red
    (earliest) to purple (latest).
    \label{fig:cascadeEvent}}
\end{figure}

Using the first year of data recorded with DeepCore, from May 2010 to
April 2011, we have observed cascades induced by atmospheric neutrinos
interacting in the DeepCore volume.  These cascades include charged
current (CC) interactions of electron neutrinos, as well as neutral
current (NC) interactions of neutrinos of all flavors.  (The
background rejection criteria used in this analysis result in an
energy threshold of around 40 GeV, so only a negligible contribution
from atmospheric muon neutrinos oscillating to tau is expected.)
Previous searches for neutrino-induced cascades in AMANDA and IceCube
\cite{Ahrens:2002wz, Ackermann:2004zw, Achterberg:2007qy,
  Abbasi:2011zz, Abbasi:2011ui} have focused on higher energies, to
avoid the background of bremsstrahlung produced by atmospheric muons.
In this analysis, we instead rely on the active veto provided by
IceCube to reduce the background of penetrating muons, and exploit the
high flux of atmospheric neutrinos at energies of a few hundred GeV to
observe a set of 1,029 cascade-like neutrino events in 281 days of the
2010 data run.  One such event is shown in
Fig.~\ref{fig:cascadeEvent}.

For this data set, recorded with the incomplete 79-string configuration of
IceCube, the smaller DeepCore fiducial volume shown in
Fig.~\ref{fig:layout} was used.  This initial configuration consisted
of only the central seven standard strings, plus 6 additional DeepCore
strings.  Based on Monte Carlo simulations, we estimate
that approximately 60\% of the 1,029 events in the final sample are
truly neutrino-induced cascades, while around 40\% are in fact
$\nu_\mu$ CC events with muon tracks too short to be distinguished in
the current analysis; efforts to further reduce this background are
underway.  The level of background due to atmospheric muons is still
under investigation but appears to be small.  The rates of observed
neutrinos are consistent with simulations of atmospheric neutrinos
using the leading atmospheric neutrino flux models from the Bartol and
Honda groups, although we are still in the process of assessing our
systematic uncertainties.  It should be noted that the predictions
based on the two atmospheric flux models differ for this event set by
approximately 10\%, due mainly to the modeling of production of higher
energy electron neutrinos by kaons.

Work is in progress to lower the energy threshold of the analysis,
which would permit observation of neutrino oscillations using the
atmospheric neutrino flux.  For baselines comparable to the Earth's
diameter, the first maximum of the $\nu_\mu \rightarrow \nu_\tau$
oscillation probability occurs at approximately 25 GeV, well within
the energy range accessible to DeepCore \cite{Mena:2008rh}. 

\section{Searches for Dark Matter}

In addition to studies of atmospheric neutrinos, DeepCore's reduced
energy threshold facilitates indirect searches for evidence of dark
matter using IceCube.  Searches are underway for neutrinos produced in
the annihilation or decay or dark matter captured in the gravitational
potential wells of the Earth, Sun \cite{Abbasi:2009uz, Abbasi:2009vg}, and Galaxy
\cite{Abbasi:2011eq}.  Because the WIMP mass must be relatively low
compared to the energy range of IceCube, additional sensitivity to
lower energy neutrinos substantially extends IceCube's reach,
especially for the lower part of the allowed WIMP mass range or for
models where the neutrino spectrum produced is relatively soft.

\begin{figure}
  \includegraphics[width=\columnwidth]{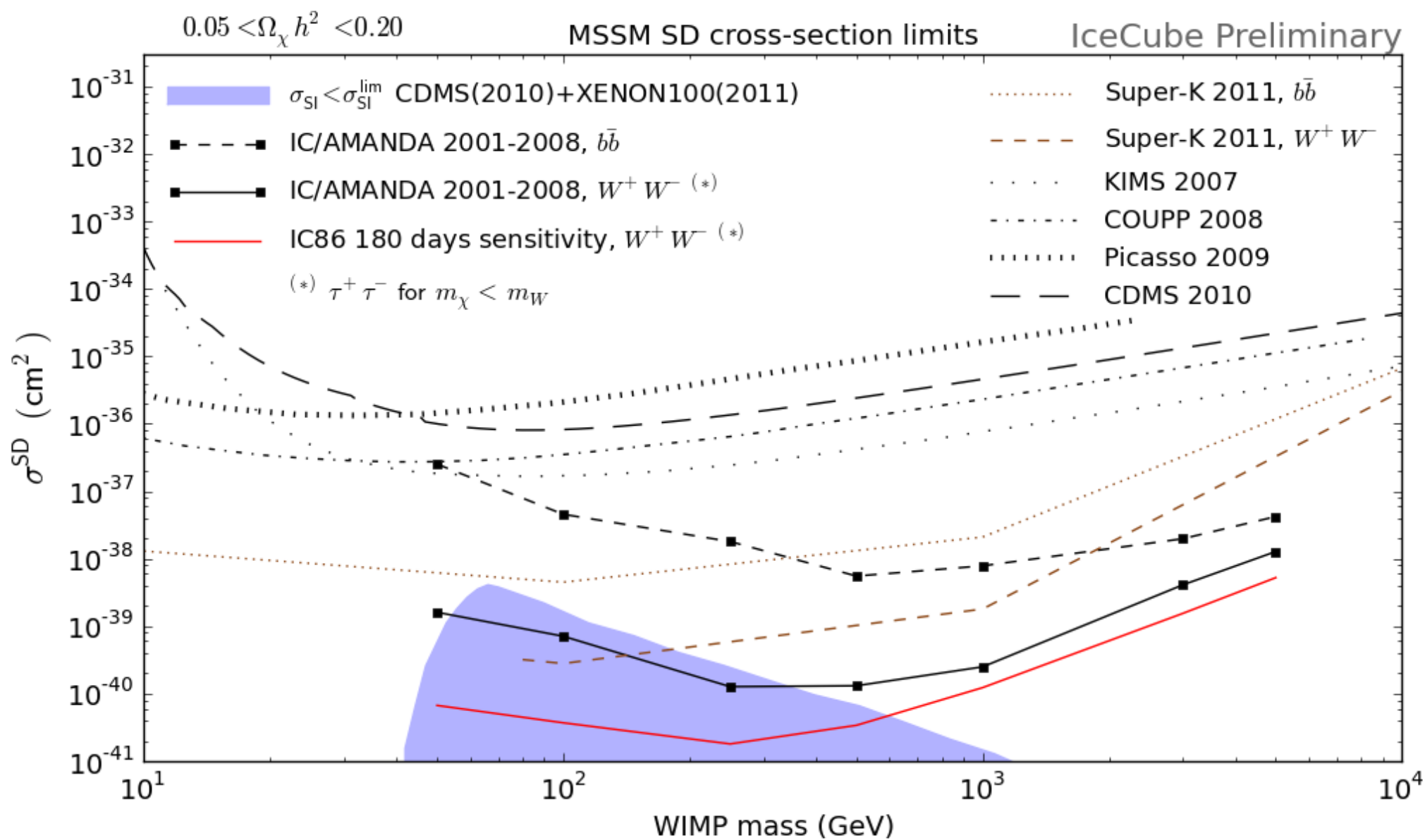}
  \caption{Limits on the spin-dependent WIMP-nucleon scattering cross
    section from various direct and indirect search experiments, and
    the projected sensitivity of IceCube with DeepCore for a ``hard''
    neutrino spectrum arising from neutralino annihilation in the
    Sun.  The shaded region indicates the possible cross sections in
    supersymmetric models not already ruled out by direct detection
    experiments' limits on the spin-independent cross section.
    \label{fig:WIMPs}}
\end{figure}

The potential of IceCube including DeepCore for detecting evidence of
dark matter annihilation in the Sun is shown in Fig.~\ref{fig:WIMPs}.
The shaded region indicates the allowed MSSM parameter space, for
models where the WIMP is a neutralino.  Direct detection experiments
have already probed substantial parts of the allowed supersymmetric
parameter space, primarily in regions where there is a substantial
spin-independent neutralino-nucleon scattering cross section, so that
coherent scattering from heavy nuclei in the detector target enhances
the cross section considerably.  For models in which the scattering
cross section is primarily spin-dependent, indirect searches
exploiting the Sun's mass as a scattering target have an advantage,
although the results depend on the branching ratios for
neutralino-neutralino annihilation channels.   For WIMP masses below
roughly 100 GeV, DeepCore provides the bulk of the sensitivity to the
neutrinos arising from Solar neutralino annihilation.

\section{Future Prospects: PINGU}\label{sec:PINGU}

\begin{figure}
  \includegraphics[width=\columnwidth]{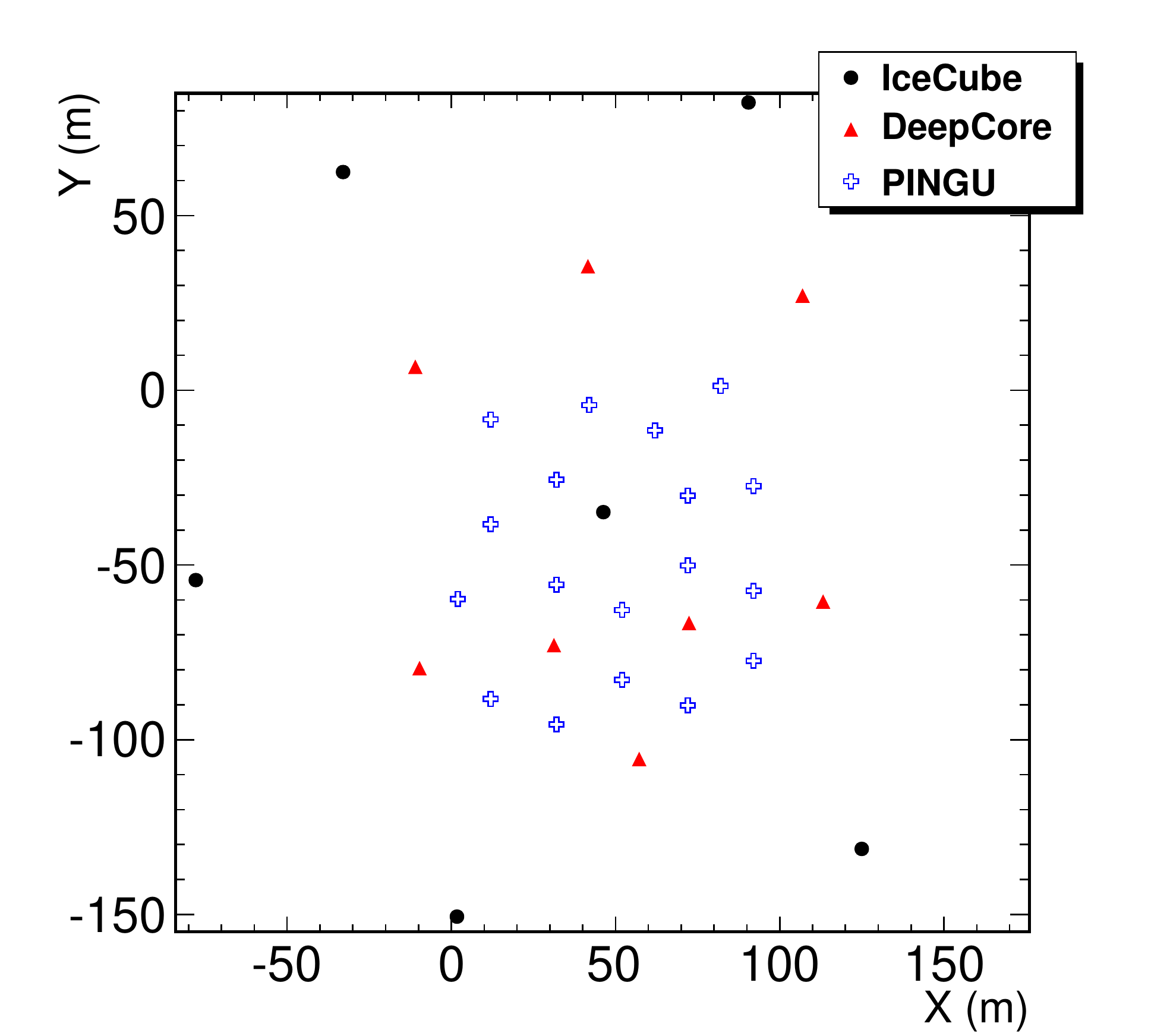}
  \caption{Top view of one PINGU configuration now under study,
    including 16 strings of DeepCore-like instrumentation.  Additional
    strings of prototype next-generation instrumentation are
    envisioned but not shown.  This layout would significantly improve
    the effectiveness of DeepCore at energies below a few 10's of GeV.
    \label{fig:PINGUlayout}}
\end{figure}

Encouraged by the initial success of DeepCore, the IceCube
collaboration and other participants are developing a proposal for a
Phased IceCube Next Generation Upgrade (PINGU), an extension of
IceCube and DeepCore which would further increase the density of
instrumentation in the central volume and further reduce the energy
threshold.  The proposal would augment DeepCore with perhaps 18 to 20
additional strings, of which the majority would be similar to those in
DeepCore.  Several strings might also include specialized prototypes of novel
sensors, perhaps similar to those incorporating a number of 3'' PMTs
rather than a single 10'' PMT, now being developed for the proposed
KM3NeT detector.  One layout of the additional strings under
discussion is shown in Fig.~\ref{fig:PINGUlayout}.   

\begin{figure}
  \includegraphics[width=\columnwidth]{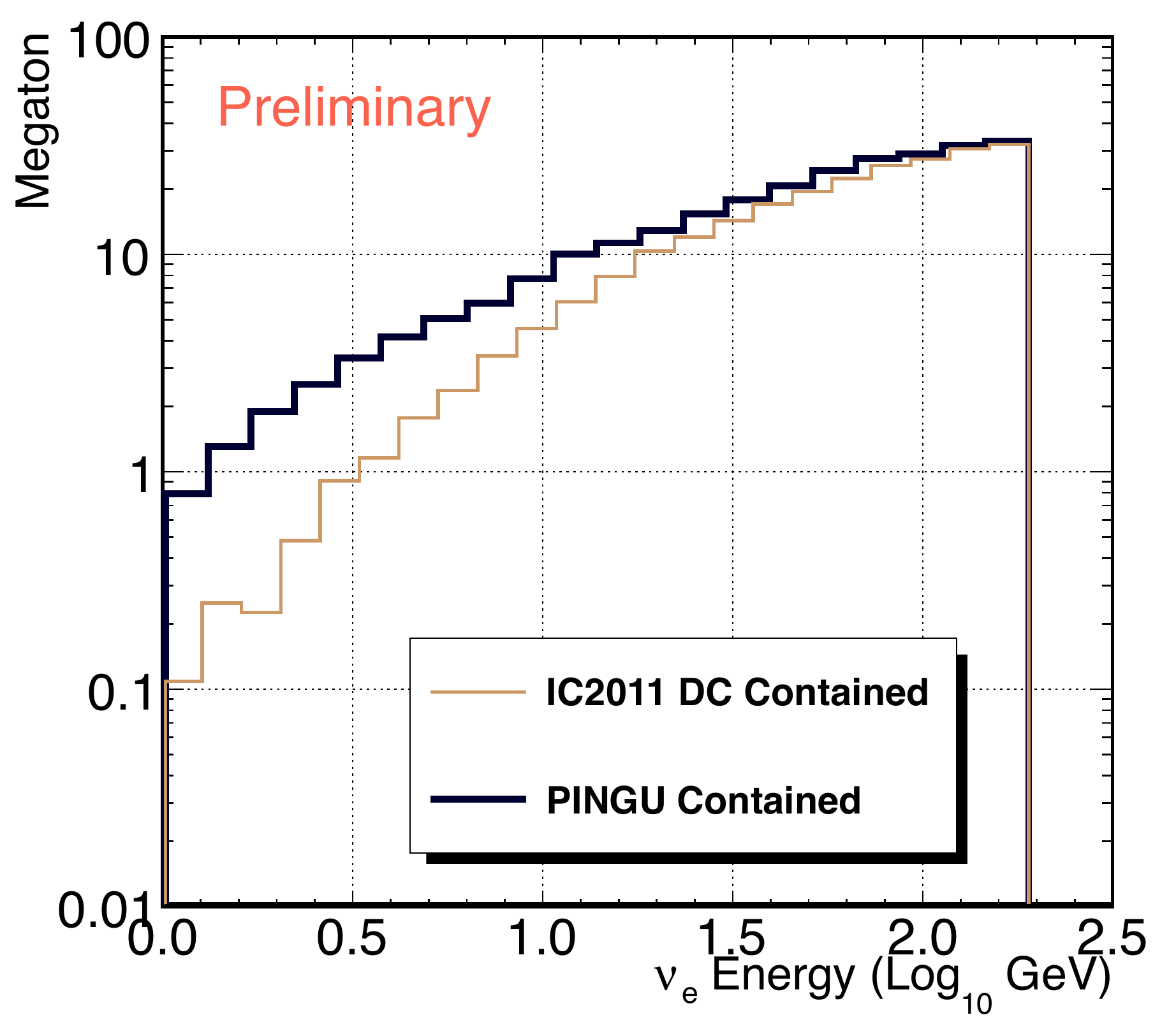}
  \caption{Preliminary estimate of the effective volume of PINGU for
    electron neutrino events at trigger level, as compared to that of
    the completed DeepCore configuration.  PINGU would retain considerable effective
    volume down to energies as low as a few GeV.  Analysis and
    reconstruction efficiencies are not included.  The geometry used for this
    estimate is similar to that shown in Figure \ref{fig:PINGUlayout} but with a slightly
    larger mean spacing between strings, so the effective volume at
    the lowest energies may be underestimated. 
    \label{fig:PINGUnuEVolume}}
\end{figure}

Such an extension would considerably increase the effective volume of
DeepCore at energies below about 30 GeV, with the potential to detect
neutrinos as low as a few GeV.  The effective volume of the detector
for events contained within the geometrical volume is shown in
Fig.~\ref{fig:PINGUnuEVolume},
as compared with that of the existing DeepCore detector.  Improvements of
nearly an order of magnitude can be seen for low energy neutrinos.
These effective volumes do not include efficiency losses due to event
reconstruction and analysis criteria, which will reduce the effective
volume achievable in final physics analysis.

\section{Summary}

The effectiveness of IceCube at energies below 100 GeV has been
significantly enhanced by the addition of DeepCore, which extends
IceCube's reach to energies of 10's of GeV.  This range is of interest
for observations of neutrino oscillations, as well as searches for
dark matter.  As a first step toward these studies, we have observed a
significant sample of atmospheric neutrino-induced cascades, enabled
by the ability of the IceCube detector to identify and veto
atmospheric muons penetrating to the DeepCore volume.  We are also
investigating the potential for a further reduction in the energy
threshold of IceCube with an additional extension known as PINGU,
which could extend IceCube's reach to energies as low as a few GeV

\end{document}